\documentclass{sigchi}

\toappear{This work is licensed under the Creative Commons Attribution-NonCommercial-NoDerivatives 4.0 International License. To view a copy of this license, visit http://creativecommons.org/licenses/by-nc-nd/4.0/ or send a letter to Creative Commons, PO Box 1866, Mountain View, CA 94042, USA.}
\usepackage{balance}       % to better equalize the last page
\usepackage{graphics}      % for EPS, load graphicx instead 
\usepackage[T1]{fontenc}   % for umlauts and other diaeresis
\usepackage{txfonts}
\usepackage{mathptmx}
\usepackage[pdflang={en-US},pdftex]{hyperref}
\usepackage{color}
\usepackage{booktabs}
\usepackage{textcomp}

\usepackage{microtype}        % Improved Tracking and Kerning
\usepackage[all]{hypcap}    % Fixes bug in hyperref caption linking
\usepackage{ccicons}          % Cite your images correctly!
\def\plaintitle{Cybernetic Health}

\def\emptyauthor{}
\def\plainkeywords{cybernetics; diabetes; quantified-self; objective-self; precision medicine; predictive health; prevention; event mining; context awareness; mobile health; persuasion; personalized}

\makeatletter
\def\url@leostyle{%
  \@ifundefined{selectfont}{
    \def\UrlFont{\sf}
  }{
    \def\UrlFont{\small\bf\ttfamily}
  }}
\makeatother
\def\pprw{8.5in}
\def\pprh{11in}

\definecolor{linkColor}{RGB}{6,125,233}
\hypersetup{%
  pdftitle={\plaintitle},
  pdfauthor={\emptyauthor},
  pdfkeywords={\plainkeywords},
  pdfdisplaydoctitle=true, % For Accessibility
  bookmarksnumbered,
  pdfstartview={FitH},
  colorlinks,
  citecolor=black,
  filecolor=black,
  linkcolor=black,
  urlcolor=linkColor,
  breaklinks=true,
  hypertexnames=false
}

\begin{document}

\title{\plaintitle}

\numberofauthors{4}
\author{Nitish Nag, Vaibhav Pandey, Hyungik Oh, Ramesh Jain\\
    \affaddr{Department of Computer Science, University of California, Irvine / Irvine, California, 92617, USA}\\
    \affaddr{Prepared March 2017, Arxiv Version Accepted 21 May 2017}\\
    \email{Corresponding Author: N. Nag, nagn@uci.edu}\\
}
\maketitle

\begin{abstract}
Future health ecosystems demand the integration of emerging data technology with an increased focus on preventive medicine. Cybernetics extracts the full potential of data to serve the spectrum of health care, from acute to chronic problems. Building actionable cybernetic navigation tools can greatly empower optimal health decisions, especially by quantifying lifestyle and environmental data. This data to decisions transformation is powered by intuitive event analysis to offer the best semantic abstraction of dynamic living systems. Achieving the goal of preventive health systems in the cybernetic model occurs through the flow of several components. From personalized models we can predict health status using perpetual sensing and data streams. Given these predictions we give precise recommendations to best suit the prediction for that individual. To enact these recommendations we use persuasive technology in order to deliver and execute targeted interventions.
\end{abstract}


\section{Introduction}
Navigation systems, like Google Maps, became rapidly popular by providing expert knowledge and real-time personalized contextual information to guide travel. Personal lifestyle guidance in such a manner does not exist in health care systems. In order to shift the focus in health systems from temporary fixes to long-term solutions such a guidance system must be implemented \cite{Sagner2016TheHealthspan} \cite{McElwaine2015SystematicClinicians}.
Commonly, physicians focus primarily on medical methods to manage health when a patient becomes ill. By evolutionary design, optimal health is universally desired (and should be provided) at all times. True health outcomes result from actions taken in every moment and place, not just medical intervention during sickness. Future advancements in health must continuously sense individual needs and rapidly provide the relevant resources so corrective actions ensure health stability.
For example, optimal health for chronic diseases like type 2 diabetes (T2D) remains a challenge. Insulin resistance, obesity and other biological changes that lead to uncontrolled T2D start many years before a formal diagnosis and treatment plan is given. These biological changes can be reversed by lifestyle choices. If insulin resistance is caught early in a prediabetic state, the course of the disease is potentially reversible \cite{Perreault2014ApproachingPre-diabetes}. 
Although biological scientific understanding has greatly progressed in the past few decades, diabetes and other lifestyle associated diseases continue to rapidly rise across the globe. Given this progress in research, we would expect the opposite. Historically, monitoring lifestyle factors has been difficult, and computational power with effective methods to address the needs of each individual have been limited. Trying to produce changes in routine lifestyle habits is also a tremendous psychological hurdle.

\begin{figure}
\centering
\includegraphics[width=0.96\columnwidth]{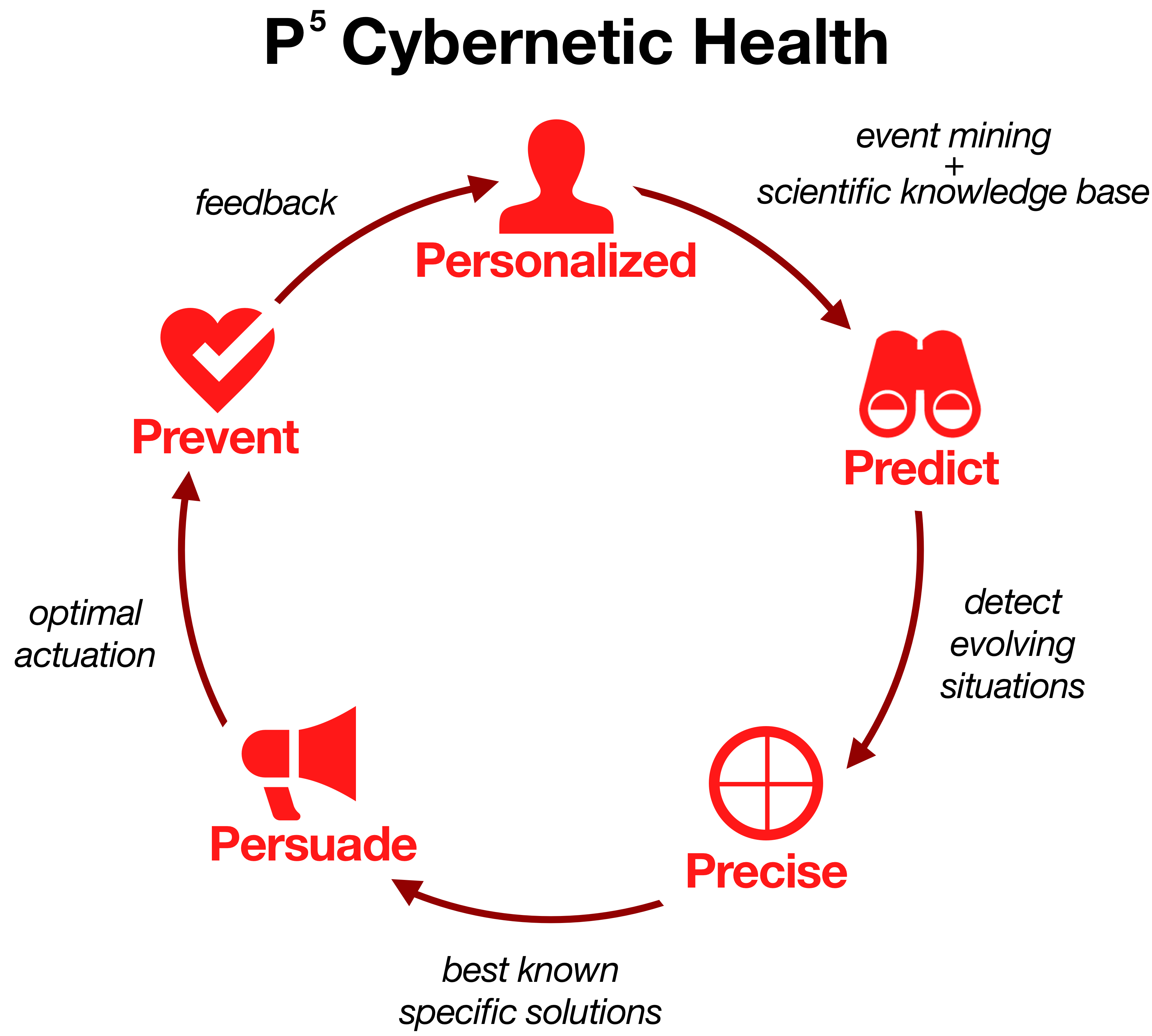}
\caption{P5 Cybernetic Health coordinates the elements of personalized, predictive and precision medicine through persuasion techniques that result in disease prevention.}
~\label{fig:p5}
\end{figure}

\begin{figure}
\centering
\includegraphics[width=0.99\columnwidth]{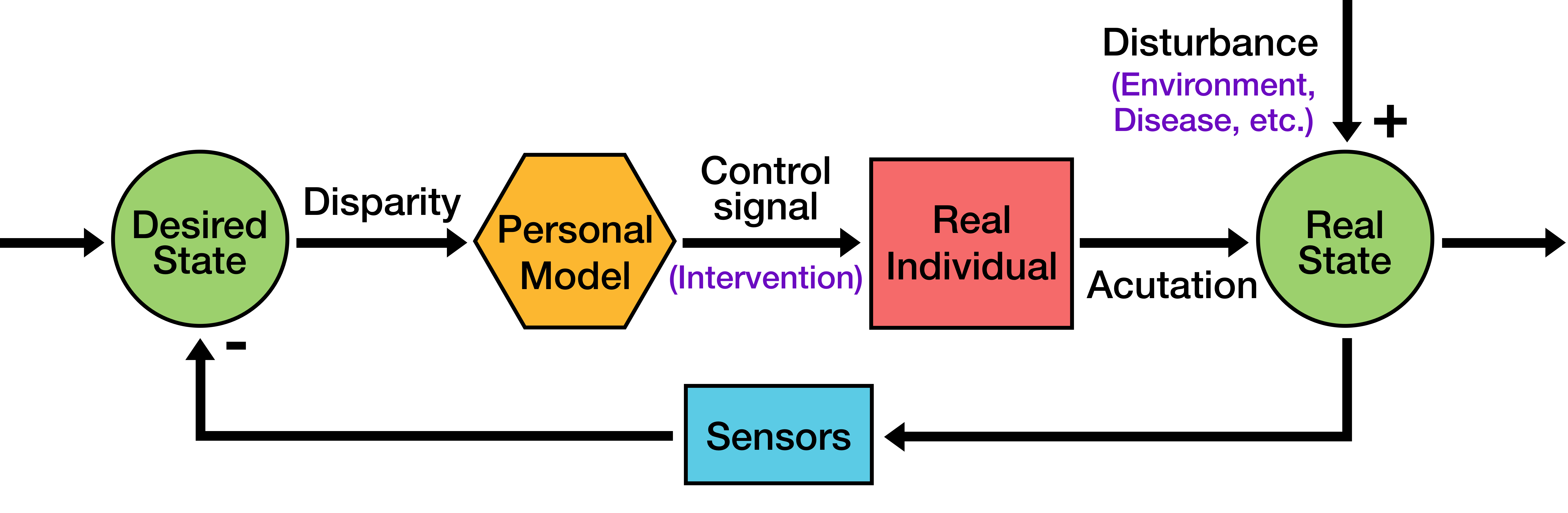}
\caption{Cybernetic Control pairs the individual user and digital health assistance to enact real-world changes to optimize health}~\label{fig:Cybernetics}
\end{figure}

This work targets three significant problems in current health care delivery through a cybernetic approach. First: Health systems largely react to problems, rather than avoiding problems through preventive measures. Second: Gold-standard medical practices depend on evidence-based medicine taken from population averages. A lack of individual contextual analysis results in compromised care and sub-optimal outcomes. Third: Access to medical guidance is limited due to poor information dissemination, and restricted physical time and space. If doctors give lifestyle suggestions to patients, they are hard to translate into everyday life decisions. When a need or question arises for health advice, such as "What should I eat?" or "Should I take this medicine now or later?", there is a large time delay to receive meaningful assistance. Patients usually scramble for unreliable information via web search engines when faced with such decisions. The difficulty of scaling physical systems, like hospitals and personnel, further limits high quality care. This is especially the case for under-served populations across the globe.

P5 Cybernetic Health (P5C) transforms these three major hurdles into opportunities (Figure \ref{fig:p5}). First: By analyzing individual data with context, we can predict problems as they arise and give the best solutions. In the instance of diabetes, we begin to predict an increase in insulin resistance risk factors for increasing complications. This leads to actionable information for the patient. Second: We tackle the issue of traditional "evidence-based" medicine by combining sub-population and individual data dynamically into a system to give "enhanced real-time personalized evidence-based" medicine unique for each patient. For a diabetic patient, we give specific actions that would result in the best blood glucose management. Third: We reduce the delay of health advice through real-time sensors and specific feedback guidance loops, while retaining the ability to scale to millions of patients through a virtual platform.

\section{Cybernetic Principles}
Cybernetic principles transformed the design of complex systems \cite{wiener1948cybernetics}. Continuous measurements are a key component in closed loop feedback control systems. Airplanes, ovens, and other machines use these feedback loops to safely and efficiently operate. Imagine if the thermometer in an oven gave a reading once every year. How would the oven know to heat or turn off? The thermometer, heater, and other components of the oven must all be coordinated and continuously working for the machine to operate correctly. Similarly, the human body maintains homeostasis amongst a remarkable array of perturbations. Biological systems use an intricate play of real-time sensors and actuators within the body to do this. Cellular sensors collect personalized information based on the individual body or tissue changes. These signals affect outputs for corrective action, and are especially effective if early warning signs are detected. These cybernetic mechanisms keep the human body naturally stable. 

Occasionally, this human system becomes unstable which results in deteriorating health. With current day medical practice, the detection and corrective actions are greatly delayed, resulting in further unstable systems and further deteriorating health for the patient. Our vision is to reduce this latency of the health care system, by using continuous sensors that are specific to the individual, while enabling corrective actions to transform an unstable health condition back to full health stability. In the operating room, anesthesiologists are beginning to develop closed-loop drug administration systems that will replace much of their own job during a surgery \cite{Rinehart2015Closed-LoopCare}. 

%Chang2015AutomationControl
For type 1 diabetics, mechanical devices can replace the biological pancreas with a hormone pump and continuous glucose monitoring. Both of these examples are mechanical implementations of cybernetics for health. Unique to our platform, we begin to mesh real world information into a virtual environment that patients can use to peek into their live health status with persuasive guidance for optimal decision making. We call an application of this Health-Butler (HB).
%augmented reality?

We chose to implement HB for T2D for two key reasons. First, diabetes is an increasing global health problem, with almost double the disease burden in the last 30 years that reaches 8.5 percent of global population. It is a leading cause of blindness, kidney failure, heart attacks, stroke and lower limb amputation \cite{diabetes}. Second, T2D and the resulting complications are largely preventable if corrective action is taken promptly. These corrective measures are actionable by individuals through lifestyle factors such as nutrition, physical activity, stress management, and environmental exposure. By continuously monitoring the aforementioned factors, we can predict if the human system will start deviating from the homeostatic set point, allowing HB to intervene at the right time and place.
     
\subsection{P5 Cybernetic Health Concept}
Leading medical professionals have advocated for integrating more technology into health care \cite{topol-2}. Future health care systems will intertwine lifestyle data with medical knowledge to develop a new paradigm that optimizes individual health. We have developed a 5 component system to bring this vision to reality. 
 
First, we use a multi-layer modeling system to understand how to build an increasingly accurate personalized model. Second, using this dynamic model connected with real time sensors allows us to predict evolving situations that an individual may encounter. Third, we use the predictions in conjunction with validated expert medical knowledge to give the most precise solutions to avoid emerging problems. Fourth, we effectively persuade the individual as an actuator in the system by optimizing their preferences, convenience, and specific health needs. Fifth, we give feedback on how the patient's actions have quantitatively affected their health. The realization of this personalized, predictive, precise, persuasive, and preventive system depends upon the coordination of available and future technologies. We call this above approach P5C (Figure \ref{fig:p5}).

\section{Related work}
Clinical and medical research in personalized and preventive medicine struggles to gain traction. Patients continue to receive sparse feedback on how to best face disease burdens in their unique daily life circumstances. Medications prevail as the primary tool to manage diabetes, because they are easy to physically scale and have standardized instructions for all patients. Currently, live exchange of personalized feedback from physical human(doctor) to human(patient) on lifestyle changes, although beneficial, is extremely costly in time and resources.\cite{Schmitt2016}

There is a significant demand for this type of virtual platform. Patients have a much higher probability of making better lifestyle choices that would combat diabetes if given guidance \cite{Sherifali2016EvaluatingDiabetes}. Unfortunately, modern (2016) smartphone tracking applications have not shown any benefit in improving glucose control \cite{Porter2016TheReview}. Physicians continue to vocalize this void in matching glycemic patterns with lifestyle history \cite{Goyal2016}.

Health monitoring research has quickly grown with pervasive computing methods. There are some research groups who have focused on lifestyle monitoring of diabetic patients. Smartphones have been used for data collection (e.g. GPS, wifi, activity) to power machine learning and symbolic reasoning to recognize lifestyle activities of diabetic patients \cite{luvstrek2015recognising}. Daily life data of diabetic pregnant women has been integrated with their network of health care institutions \cite{ballegaard2008healthcare}. Other groups have focused on health-related data monitoring for chronic disease care. Waki et al. implemented a smartphone self-management system which consisted of 4 modules; 1) data transmission, 2) evaluation, 3) communication, and 4) dietary evaluation, which resulted in improved HbA1c in 3 months \cite{waki2014dialbetics}. Mukherjee et al. provided an environment for caregivers to monitor patient data in real-time \cite{mukherjee2014patient}. Katz et al. and Mamykina et al. designed mobile systems to merge and analyze data streamed from multiple sensors to give user recommendations \cite{katz2016investigating, mamykina2015adopting}. Banos et al. explored existing personalized health data applications to develop a framework, called Mining Minds, to assimilate health data in order to better serve patients \cite{Banos2015MiningSupport}. However, to the best of our knowledge, there is no joint research between medical and computing fields that cover the scope of cybernetics to coordinate the elements of personalized, predictive and precision medicine through persuasion techniques to result in disease prevention (Figure \ref{fig:p5}).

\begin{figure}
\centering
\includegraphics[width=0.89\columnwidth]{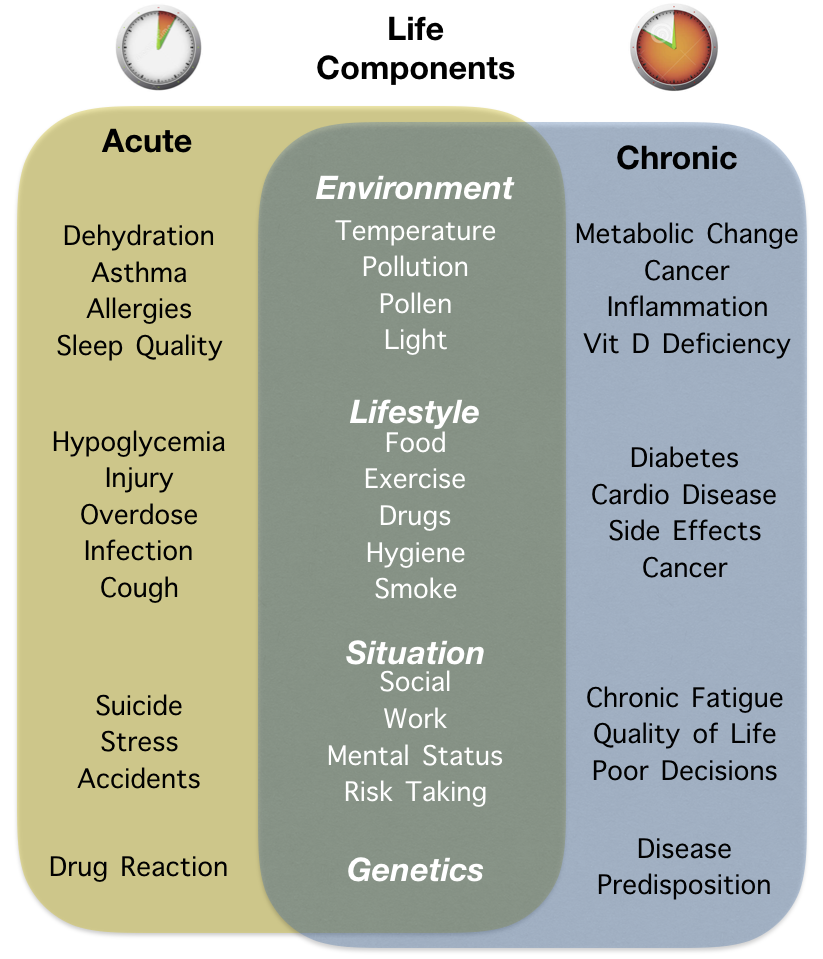}
\caption{Cybernetic systems apply in all time scales of health from acute to chronic diseases. Here are just a few examples of components that can have both acute and chronic effects on health homeostasis. Other than genetics, most of these factors are controllable to some degree.}~\label{fig:venn}
\end{figure}

Computing work in this field has primarily focused on giving the user figures and statistics of past data. This is true for both hardware and software in personal health. Hardware such as the Fitbit, and health software like Apple's HealthKit only function to acquire and accumulate data. This does not fulfill the function of providing timely and personalized health advice in a predictive manner. Most importantly current digital health mechanisms are rudimentary in detecting context for each individual. Second, recommendation engines that are used in health applications ignore mechanisms to maintain retention and trust of the user. Users quickly get alert fatigue from poor recommendations. To sustain users, applications must give users autonomy, cater to their desires and convenience, while also informing them in an encouraging manner. Additionally, many lifestyle data parameters are gathered through manual mechanisms. For example, popular nutrition tracking applications ask users to manually enter information. This further causes a high loss in user retention, while having poor accuracy of input values \cite{Krebs2015HealthSurvey}. Users desire low data entry burden, with high functionality to help reach their goals \cite{Krebs2015HealthSurvey}. 
 
\section{HealthButler Application}
\begin{figure}
\centering
\includegraphics[width=1.01\columnwidth]{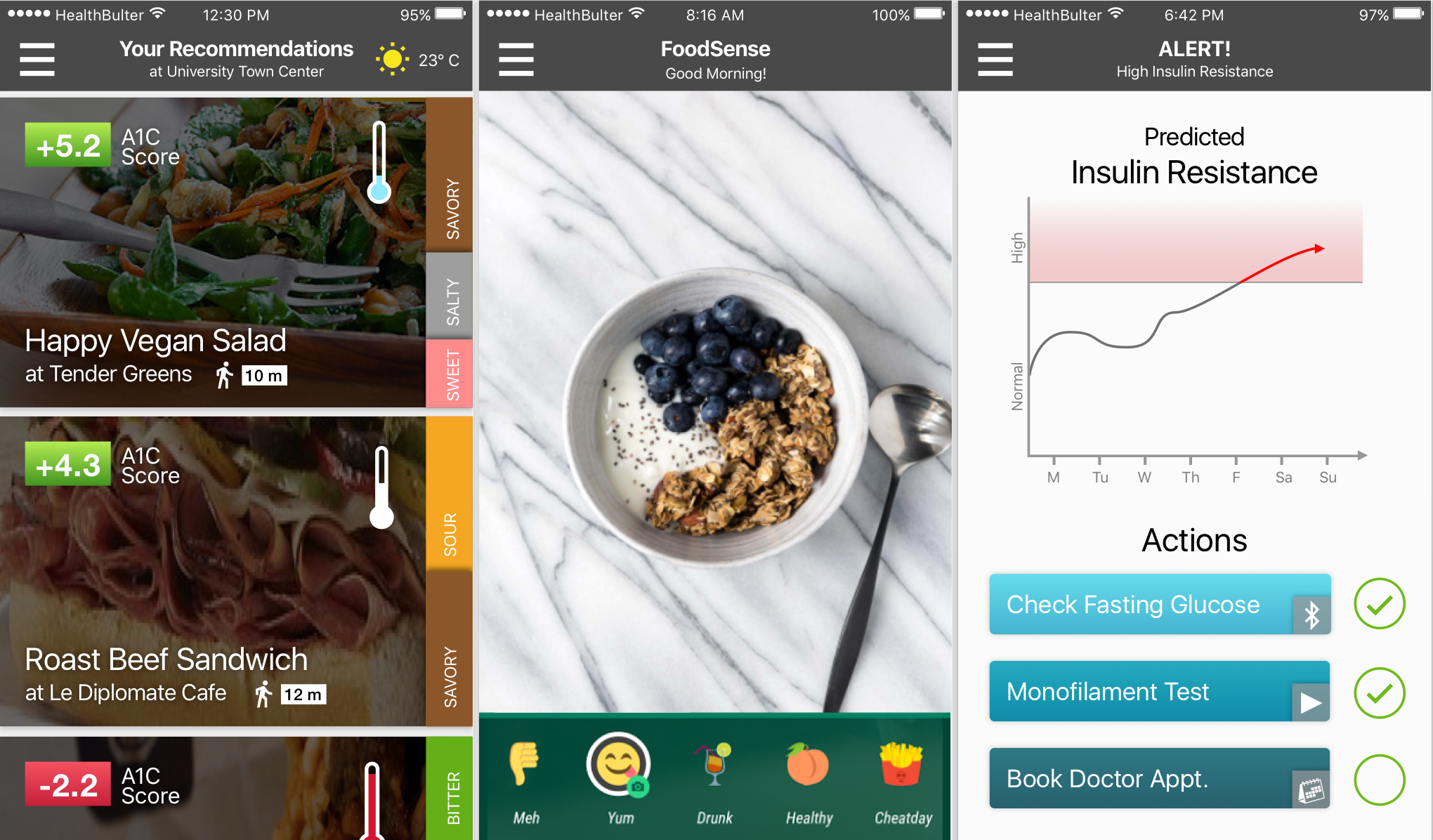}
\caption{Left: Nutrition guidance is catered to the user's preferences, needs, and available resources. Center: Easy one-touch food and mood tracking. Right: real-time health status is shown, with direct actions to take for help.}~\label{fig:screens}
\end{figure}

To illustrate the delivery of P5C through HB, we describe how a T2D patient named "Bruce Uberschweet", uses the system to optimally manage his health condition in both positive and negative scenarios. This will include analysis to better control blood sugars and reduce drug dependency through improved metabolism \cite{McGarrah2016TheAction}. Bruce is looking for lunch on his commute to work on Monday and knows that HB always gives him the fastest access point to tasty and nutritious food. An intelligent recommendation engine takes into consideration his real-time personal tastes, logistical convenience, and current health needs to provide him a curated list of specific dishes that he can easily pre-order. He can also clearly see how each dish affects his diabetes so he can feel empowered to choose what is good for himself (Figure \ref{fig:screens}). After attending a wedding on Sunday, HB predicts a rising insulin resistance based on his previous lifestyle data, and gives him immediate actions to take in order to address the worsening condition. It simplifies his next steps such as booking an urgent appointment with his doctor to change his medication dose (Figure \ref{fig:screens}). These are two examples of how HB is actively engaged in predicting Bruce's health status, merging in with his daily life in an unobtrusive and useful way. Bruce can actively see how his external world and internal body are interacting through HB.

\section{P5 Cybernetic System Framework}
\begin{figure}[t!]
\centering
\includegraphics[width=0.9999\columnwidth]{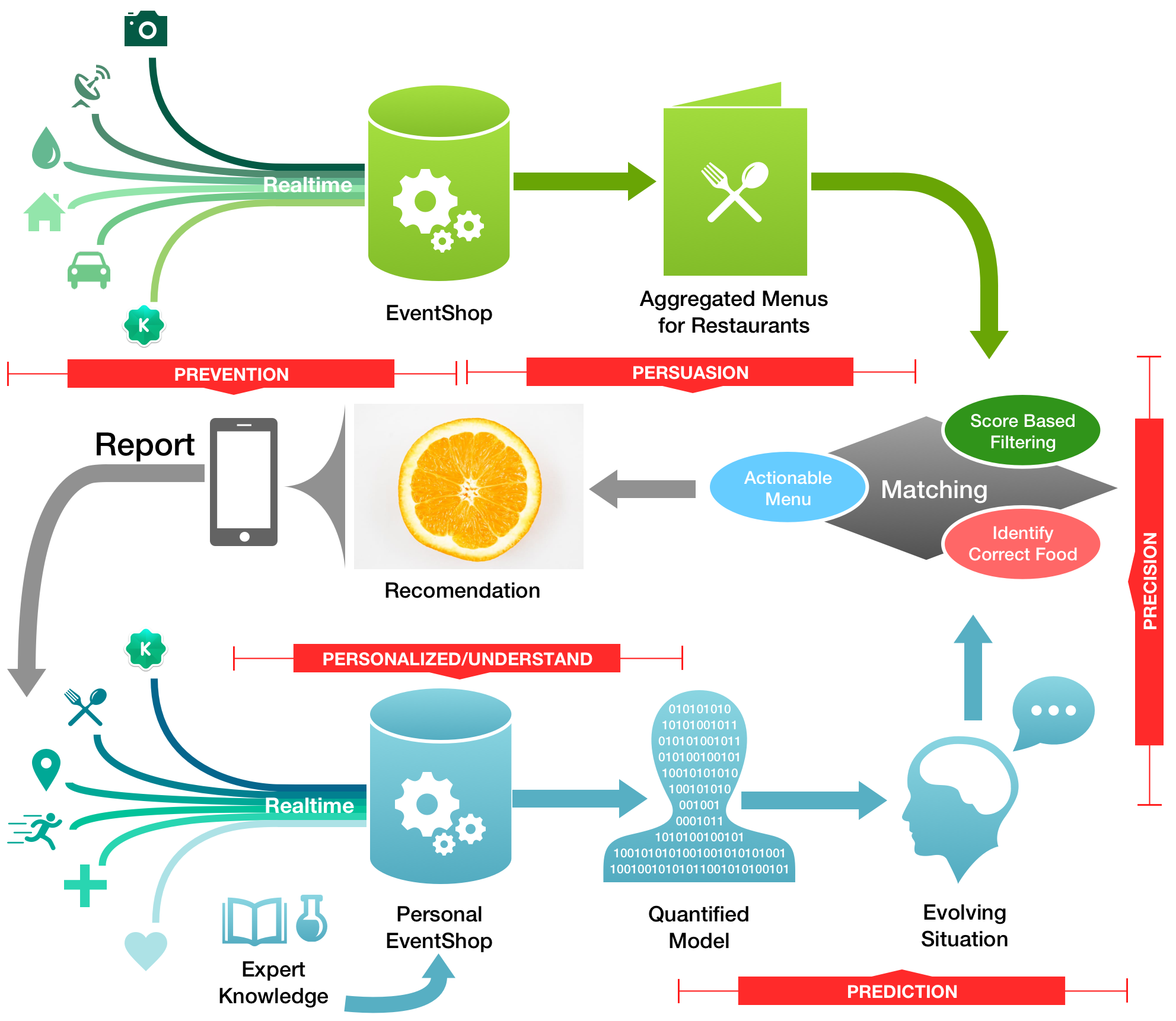}
\caption{System Architecture of P5 Cybernetic Health}~\label{fig:system}
\end{figure}

Producing the exemplary application of HB requires the coordination of multiple modules in the P5C system. Each component is integrated into the system architecture in (Figure \ref{fig:system}) to produce the front end user interface of HB.

\subsection{A. Personalization}
%(will elaborate on Lifelogger, Personal Eventshop, Objective self, Habit waveform etc).
Allopathic medicine provides solutions based on averages from large clinical trials. This method improved population health dramatically in the last century, but is now hitting a bottleneck. Various diseases are on the rise despite the latest advances in biomedical science. Physicians and researchers do not have the capacity to maintain and analyze detailed records for every individual. With the recent advances in sensors, smartphones, and pervasive technology, it is becoming possible to record data to create a digital imprint of each user,resulting in the concept of quantified self \cite{swan2013quantified}. For example, mobile applications such as Google Fit, Moves \cite{moves_app}, or Fitbit, are actively recording user life data. Dey et al. started building a conceptual framework, named Context Toolkit and AWARE, for developing quantified self applications that understand context \cite{dey2001conceptual, ferreira2015aware}. Creating individual models became the next logical step in personalized systems. For example, Objective Self (OS) began to build a comprehensive human model using heterogeneous data sources for each individual \cite{jain2014objective}.

\begin{figure}[!h]
\centering
\includegraphics[width=0.7\columnwidth]{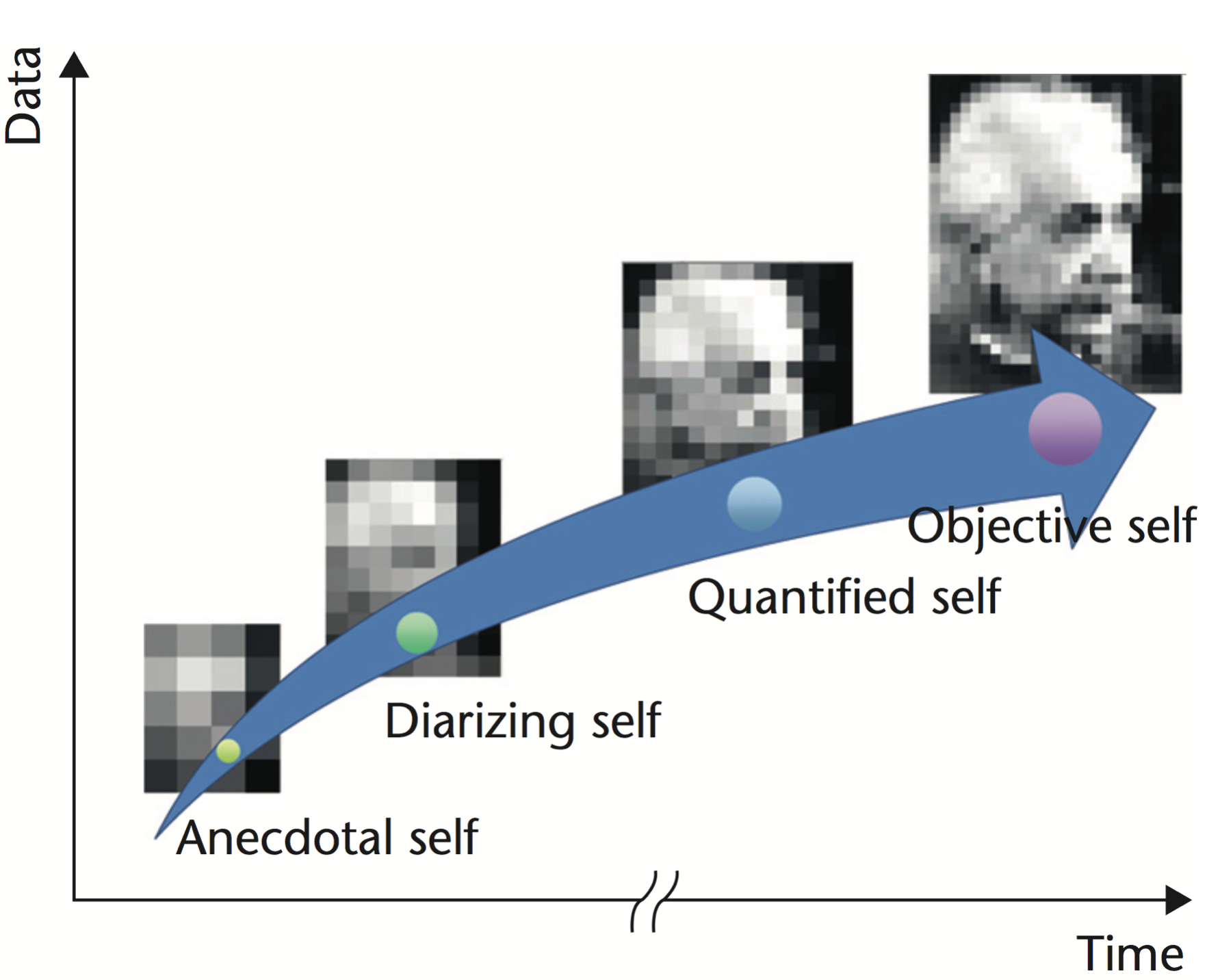}
\caption{As data availability increases over time, we are able to build more accurate models of individuals. In ancient times, anecdotal oral and diary traditions documented life. With increasingly quantitative data we can build increasingly accurate models of human life.}
\label{fig:anecdotal}\textbf{}
\end{figure}

\begin{figure}
\centering
\includegraphics[width=0.95\columnwidth]{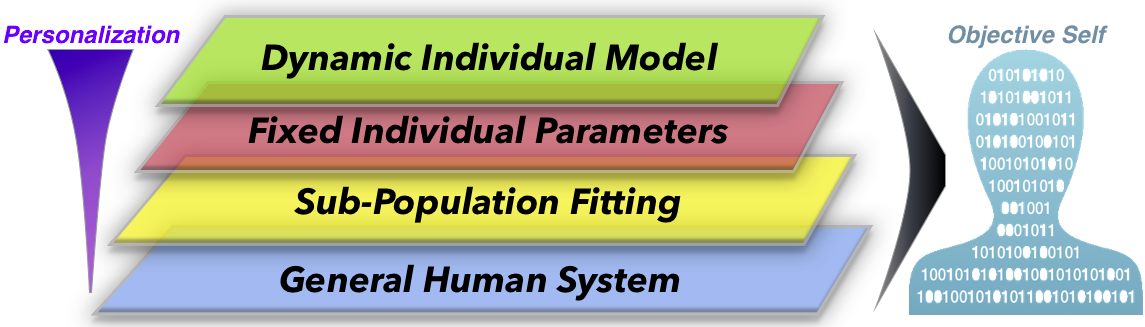}
\caption{We integrate increasingly personalized layers of data modeling to build an objective self.}~\label{fig:4layer}
\end{figure}

P5C builds upon these concepts to build a more comprehensive understanding of an individual. Smartphones can recognize many events in our daily life \cite{JordanThesis:2015, oh2015intelligent, ferreira2015aware, biegel2004framework}. The timeline of these life events that can be referred to as a personal chronicle (personicle) \cite{jalali2014personicle}, and in the near future will be gathered lifelong, from the 'womb to tomb'.
%  Objective Self
P5C is primarily focused on changing the lifestyle of a person to bring about clinically relevant positive health outcomes. Patient data is segregated into four levels of increasingly personalized rules (Figure \ref{fig:4layer}). The cybernetic system will target the optimally desired health state. In the first level, we apply universal general rules to build a skeleton individual based on medical and biological expert knowledge. Second, we incorporate specific knowledge that applies to sub-categories of people, such as, gender, ethnic background and more. This layer is applied as a function to the individual at a specific time and place. Third, we take into consideration firm variables about the person, such as genetics, age, home location, socioeconomic status etc. Fourth, we build dynamic individual models using the above three rule layers in addition to an event mining platform that ingests individual sensor data in real-time. The fourth layer captures the user's personicle based on life activities, food intake, medical and physiological parameters, emotional status and environmental conditions to build the live user OS. In the future, additional data streams can easily be incorporated into this data mining platform.

\subsubsection{ActivitySense} % by Jordan, 02/01/2017
Lifelogging through sensors makes it possible to record the totality of an individual's experiences \cite{gurrin2014lifelogging}. Personalized lifelog raw data is then extracted by HB to semantic-level activities in real-time. These real-time events allow for intuitive analysis of a dynamic human life. Life activity events for health computation are parallel to objects in a picture for intelligent visual computing. To build realistic systems, we can never completely capture everything due to the lack of standard data formats and the complexity of life \cite{sellen2010beyond}. Therefore, the range of our semantic activity also needs to be refined in some number of meaningful activities rather than recognizing all possible things. Quantitative information of time use, frequency, intensity of stress, enjoyment, and other affective states are most meaningful to medical researchers \cite{kahneman2004survey}. HB currently targets 17 standard semantic level activities which include: socializing, relaxing, prayer, eating, exercising, home events(watching TV, preparing food, sleeping, housework), shopping, conversations, computer/e-mail/Internet usage, working, commuting, and important diabetes-related activities, such as the use of the toilet and hospital visits. High blood sugar increases urination frequency in diabetic patients.

Some research groups have been working on semantic-based life event recognition. Routinely visited locations such as home, work, or school can be tracked and these can indicate pursued activities such as leisure, working, or traveling \cite{liao2007extracting}. Contextual information can be processed together to infer everyday activities on a high level such as eating, cooking, walking, or talking etc \cite{wang2012semantics}. Mobile Lifelogging tracks activity \cite{moves_app, Fitbit_app} but also detects high-level life events \cite{Life_Cycle,JordanThesis:2015}. 

\subsubsection{FoodSense}
Patients suffering from obesity, diabetes, cardiac disease and other chronic conditions continue to have difficulty following nutritional guidelines. Primary reasons include the failure to address individual differences, resources planning, and high burden of manual data entry \cite{my_fitness_pal}. A quantitative diary of food intake may be helpful in regulating dietary habits, but this type of a system is still not pervasive \cite{Darby2016}. Patients that measure food in conjunction with glucose and insulin only see a slight benefit \cite{Friedman2013}. An essential requirement is to have a personalized and objective approach that requires minimal user input.
Many companies and researchers try to encourage people to manually enter information whenever they eat. The "Accu-Chek 360" by Roche uses 7 glucose time points a day for 3 days and has been shown to only be slightly beneficial to clinicians and  in most cases, however, only partial qualitative information was reported.
FoodSense captures and analyzes photos or transactions of food to create a quantitative nutrition diary along with emotions in one touch. Using these pictures we use open deep-learning techniques by Google and Clarifai for food recognition in photos \cite{clarifai_food}. Nutritional parameters for each person are analyzed based on their personal health status \cite{Goyal2016, Wood2015a} to give them the most relevant suggestions on guidance. Bruce receives his lunch recommendations on HB from this analysis, in conjunction with all his other data (Figure \ref{fig:screens}). For example, Bruce exercised before he received his nutritional recommendations which gave him options that are suited for glycogen replenishment. When he purchases the item from HB, it is automatically recorded into the system, with no extra user burden. On a broader scale, this decision making support shifts consumers towards healthier menus, and persuades businesses to offer healthier food options. \cite{walmart, mcdonald}

\subsubsection{MoodSense}
This measures the user's mood at a particular time. Relating this to life events is vital for making effective recommendations based on what the user enjoys or dislikes (described in later sections). We get an estimate of the user's emotional state using two types of inputs: 1. Active/Explicit input allows the user to directly mark a moment or an event on their timeline with their emotional state (Figure \ref{fig:screens}). We can use this application to correlate the emotional state with events (co occurring and delayed) and if the correlation is significant then we can associate that emotional state with the events. 2. Passive/Implicit input allows us to monitor the emotional state of the user by keeping track of their various interactions with environment and people. This includes their social media content and text communication. Similar emotion detection studies show strong promise with these methods \cite{Tausczik2010,Gamon2010}.

\subsubsection{MedicalSense}
Incorporating medical data into OS is essential to make informed health decisions. For diabetic patients, we collect blood glucose values through bluetooth glucose meters, their medication compliance with bluetooth pill boxes, and their pre-existing health conditions data from the hospital electronic health records. As continuous glucose monitoring becomes technologically advanced, patients will be able to report their blood sugar values without any invasive interventions \cite{AlphabetVerilySciences}.

\subsubsection{EnviroSense} 
Environmental factors continuously affect the health of every individual. Having the ability to measure the local environment of each individual, we will give insight into how they are being affected by factors the user is otherwise unaware of. In diabetes, air and water pollution have been shown to increase the risk of diabetes \cite{Eze2016, Chen2013, Eze2014, Brauner2014}. 
Long term exposure to particulate matter in the air can activate pathophysiological responses that can induce insulin resistance \cite{Chen2013, Eze2015}. 
While public data regarding the quality of the environment is readily available, it is not incorporated and tracked at the individual level.

We are using an open source software platform, called \textit{EventShop} to ingest and assimilate different data streams \cite{pongpaichet2013eventshop,tang2015habits,singh_situation_2012}.
It combines different environmental data streams ranging from climate data, air quality, pollen counts, and micro-blogs (like Instagram and Twitter) to understand how the environment is evolving. For each individual, the environment stream will be stored by obtaining this information from EventShop. Additionally, we use available data from open-sources such as Yelp, Google Maps, and government websites to understand what resources are available to a user at any given location and time. These data sources will be used in the need-to-resource matching for daily life such as food suggestions, activity recommendations, emergency hospital directions and more.

% LifeLog
% Personal Eventshop
% Habit waveform/impulse response
We can create a behavior profile for the user with the input from the above modules, which can be referred to as \textit{habit waveform}. 

The habit waveform represents a user's behavior averaged over a large period of time, and it is continuously affected by the person's actions and environment. We can view this as a control system where the habit waveform represents a person's equilibrium or steady state and all the events can be viewed as an impulse provided to the system. Our goal is to modify the steady state over a period of time to a configuration known to represent healthier lifestyle and minimize the effects of the events/impulses which are detrimental to the health of the user.
%We need to identify the relevant features that we need to track in each of the event streams and the habit waveform will give you the aggregated value for those features based on the person's behavior. 
% Need more concrete thoughts on how we are going to formulate the problem, do we show this as a control loop with a decay function, and manage the damping as a function of the variance of input stream, that way people with variation in their habits are shown a wider range of options as recommendations.

\subsection{B. Prediction}
% (Event mining, AI/ML, Action Waveform, etc)
% Efforts made in predictive monitoring fail to integrate at a systems level \cite{Garde2016RespiratoryStudy}. 
 %This lack of integration in healthcare needs attention \cite{Lobelo2016TheReduction}. To create effective approaches in learning from data, we will develop an interactive event mining framework that integrates expert knowledge. 

We use data driven analysis and pattern mining algorithms to find event patterns in a personicle to build individual OS models. These models estimate and predict how future events will affect the habits waveform and variables of interest.
%For T2D, patients are measured over three month spans with their blood HbA1c levels. These models estimate the relevant variables for the person and can be used as a motivator for changing lifestyle without burdening the user to manually measure sugar levels.

Event relationship operators formulate compound events and compute co-occurrences which are then tested with a new set of data \cite{jalali2016interactive, jalali2015bringing, jalali2016human}. This framework extends traditional complex event processing \cite{buchmann2009complex} significantly by including space, multiple event streams, with point and interval events to enable real world data analysis.

\begin{figure}
\centering
\includegraphics[width=0.6\columnwidth]{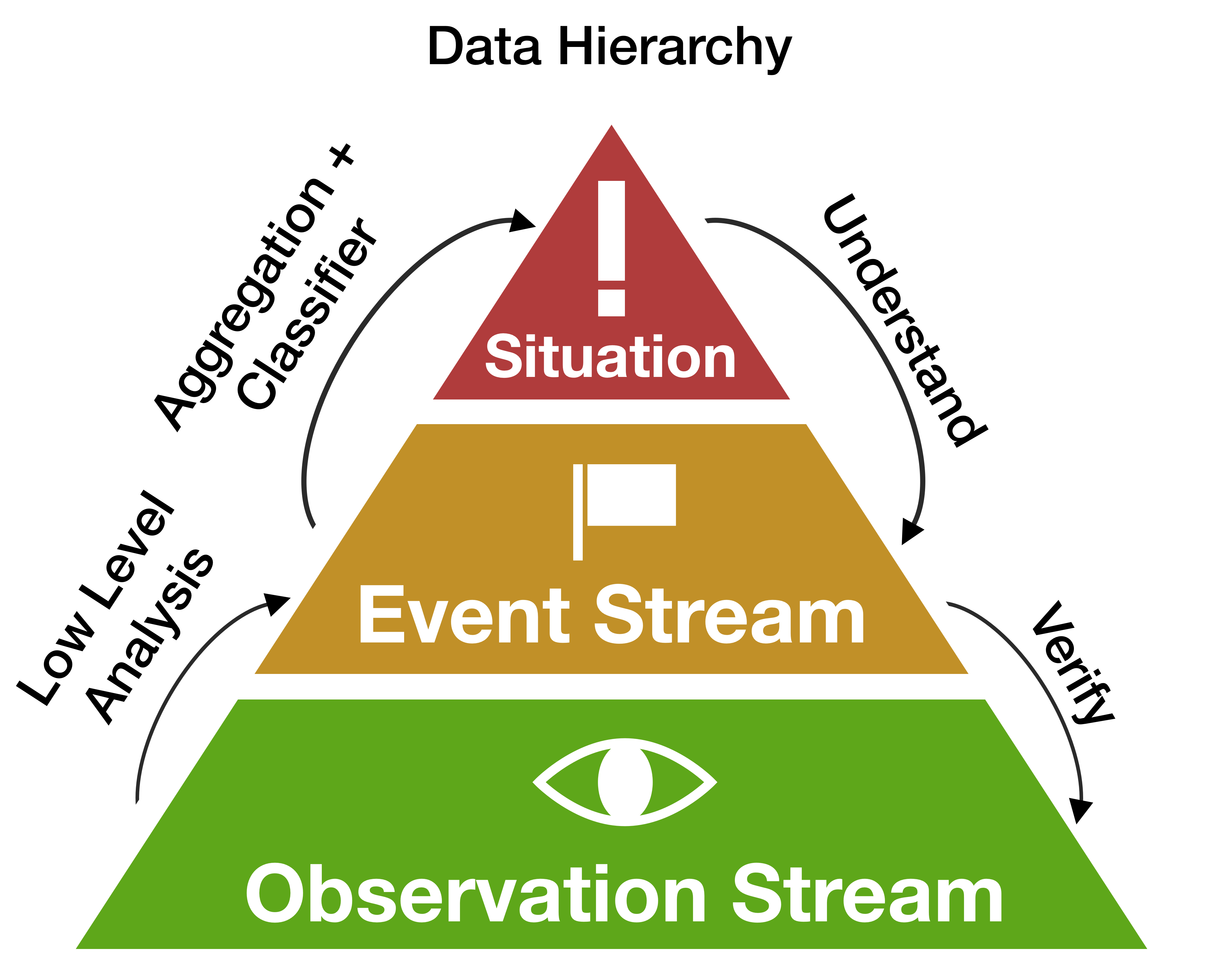}
\caption{Observations are signals gathered from sensors. Events bring semantics to the raw data. Situations give a cognitive understanding of the current and past states. We use this to predict future states of the system.}
~\label{fig:data_pyramid}
\end{figure}

In using the event based computational paradigm, our analysis follows very intuitively from raw data to events to situations, which can be directly related to physiological measurements (Figure \ref{fig:data_pyramid}). Thus we are extracting live raw data to find a direct relationship between events of the user and their physiological condition (situations). This allows for clinically valid interpretations that feed into the recommendation engine.

Merged event streams in a personicle \cite{jalali2014personicle} produce a stream of time-indexed events, e.g., high fat meal eaten at 3pm Monday or 40 minutes of exercise at 2pm on Saturday or high blood glucose level at 5pm Thursday. Statistical models \cite{heins2014statistical} are used to identify recurring patterns in a sequence of events. By fitting such models it is possible to identify sequences of events that may predict high likelihood of an adverse medical event. Fitting the model to an individual's event stream data is a challenge that may require weeks of observations. Bayesian hierarchical models \cite{gelman2014bayesian} can be used to leverage information from a population of users to give upfront meaningful analysis until the data from a single user is sufficient. This approach provides an intuitive data-determined degree of synergistic sharing between individual and population information. Parameters of models that fit separate individuals can be described by a population distribution where recurring patterns are shared while some remain unique to each individual \cite{heins2014statistical}. 

Essentially, our system initially provides strong population data driven assistance while becoming increasingly personalized as data accumulates.
We can also use these event patterns to identify the food preferences of a person. Clustering and factor analysis are used to identify eating habits using empirical methods\cite{Newby2004} in addition to using defined diet quality scores\cite{Waijers2007}. In the example of diet, once we have identified the preferences that are relevant, we create the personalized food habit waveform to illustrate the degree to which a user consumes particular food groups. Similar analysis is done for activities as well which gives us the historical habits of the person. 
Life habits coupled with the MoodSense data identifies preferred activities. For example, the person may be stuck in traffic everyday while commuting to work (a life habit), but using the emotional response we can ascertain they do not prefer waiting in traffic. User preference prediction identifies activities/food items which are most likely to be executed. We focus this prediction to trigger positive changes in the user's health. As we will see in the persuasive aspect of the system, food events/activities which have a positive impact on the user's health and satisfy the above constraints are the most appropriate suggestions for the user.

Aside from this, we can use the event history of the user to identify anomalies in their behavior. Some of these anomalies may represent medically significant behavior changes, for example if we can identify sudden mood shifts using MoodSense, there is a high likelyhood of hyper or hypoglycemia. This activates the system to provide emergency relief services. We also use the personalized data to predict developing insulin resistance over time. An accumulation of low activity, high fat and sugar foods, with associated lethargy indicates the individual is not on track to improve their glucose sensitivity. After Bruce's attended the wedding on Saturday, a combination of these factors trigger the alert and actions based on a predicted increase in insulin resistance (indicated a deteriorating diabetic condition) (Figure \ref{fig:screens}).
%include rule based systems
%4 layers, where what fits, sources and models for each layer
% how an activity affects the relevant variable/habits waveform and use that to choose the best possible course of action
%use context of the person/environment while making the recommendations
%to predict:
% A1c
% Insulin resistance
% most likely next behaviour
% most probable followed suggestion
% anomalies in behavior
% adverse effects/ emergencies
%

% Using the personicle OS, a healthcare professional or researcher can analyze an individual, have a knowledge-based system that can adjust the lifestyle of that individual \cite{jalali2014personicle}, and understand disease models as a whole \cite{jalali2015complex}.

\subsection{C. Precision}
A lack of data on personal lifestyle in relation to biomarkers has been a struggle in the quest to provide the most precise treatments for patients \cite{Valencia2017}. President Barack Obama also began the Precision Medicine Initiative to begin following various cohorts of patients to understand what constitutes better treatments for some over others \cite{Fradkin2016NIHResearch}. Researchers are also trying to link genetic factors to diabetes outcomes, but they are confounded in their research due to a lack of high fidelity lifestyle data \cite{Type2DiabetesGeneticsTypeGenetics}.

By predicting the probability of physiologic events, we can send the most appropriate control signals to the user to take corrective action to maintain their health status before it starts becoming unstable. We develop an algorithm that incorporates various factors such as the severity of the adverse event along with the likelihood of occurrence. To prevent alarm fatigue, we dispatch the control signal only after the threshold of maintaining optimal health is crossed. 
%Otherwise the control signal of actions are given as choices to help guide the user towards better health goals. 
Precise diagnostic tools, medications and other medical interventions are also suggested to the physician as to reduce waste of resources and ensure better outcomes. Most importantly, giving the most precise treatment for an individual relies on the generation of actionable interventions \cite{Valencia2017}. To produce effective changes in diabetes status, glycemic patterns need to be accurately related to a continuously monitored lifestyle \cite{Goyal2016}. This allows for a clear understanding of how different factors affect a patient's blood glucose. Early lifestyle corrections are a primary method to prevent microvascular complications of diabetes, but include potential non-intuitive actions such as having a moderate amount of alcohol in the diet \cite{Valencia2017} or switching to a flexitarian diet (reduced meat intake) \cite{Derbyshire2017FlexitarianLiterature}. Specific exercises improve insulin resistance over others, especially aerobic training over resistance training \cite{Valencia2017} \cite{Marson2016EffectsMeta-analysis}. Certain patients are at higher risk of hypoglycemia from medications, and thus may have a less stringent glucose target to prevent severe hypoglycemic episodes, while also needing to be more informed of potential factors that may cause hypoglycemia \cite{Yun2016RiskMellitus}. Furthermore, exercising to lose weight is not necessarily the best therapy for poor glucose control \cite{Franz2015LifestyleTrials}. 
Some clinical testing methods, such as monofilament testing, to check for diabetic peripheral neuropathy can also be easily executed by family or friends near to the patient. P5C uses a host of verified medical data in conjunction with physicians to direct precise diagnostic, treatment or control actions \cite{UpToDateUpToDate}. HB will suggest these actions in addition to prompting a doctors visit when predicted glucose control and insulin resistance from lifestyle data is not in the normal range (Figure \ref{fig:screens}. We deliver these non-intuitive signals to the user to take action via HB. Ultimately, merging individual quantified models with expert knowledge, translation of medical research to help the unique case of each individual is accelerated.

\subsection{D. Persuasion}
\begin{figure}[t!]
\centering
\includegraphics[width=0.7\columnwidth]{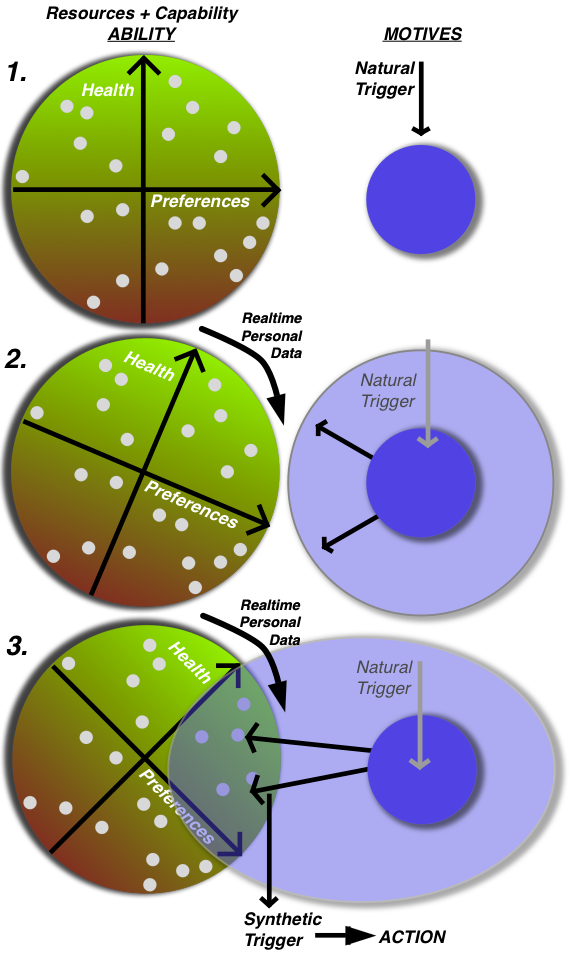}
\caption{ 1. The upper right quadrant of Ability represents options that are both healthy and preferred for a given user. Natural triggers (such as hunger) cause an increase in motivation to look for food. 2. As the motivation increases, the user mobilizes to access resources to fulfill the motivation, generally aligned with their preferences. HB tilts the Ability to optimize for convenience, preferences and health factors in real-time. 3. As the user is presented with choices, we use a synthetic trigger to increase the probability of the optimized action.}~\label{fig:persuasion}
\end{figure}

%  include persuasive theory citations
%  introduce motivation, ability and trigger, and relate the corresponding elements in our case
% cite importance of persuasion coupled with logging
A 2016 review of modern smart phone food tracking has not shown benefits for glucose control \cite{Porter2016TheReview}. This highlights the need for considering the user's preferences while making recommendations. The incentives also need to be aligned for the patient to take positive actions \cite{Goyal2016}. The goal of our system is to induce gradual habit changes via suggestions which cater to user's preferences and cause incremental improvements in their long term habits and health.
%If such actions are repeated by the user then we are improving their health in the long term. 
Recommendations are a cost function of preferences along with health impact. Healthy options which are diametrically opposite to the person's preferences have a high cost as the user is unlikely to act on the suggestions. Thus we align their preferences by inducing minimal changes which are good for their health (Figure \ref{fig:persuasion}). Concepts from persuasive technology help us in generating recommendations that are most suited for the individual\cite{Fritz2014, Arteaga2009}. According to Fogg's Behavior Model \cite{Fogg2009}, there are three factors which determine behavior: motivation, ability and trigger.

\begin{description}
\item[Motivation] refers to the individual's willingness to follow through with the suggestion. There are three factors which can affect motivation for any activity/event:
% \begin{itemize}
% \item 

\textbf{Preference/Instant gratification} motivates the events which the user enjoys and would always prefer to perform (if the other factors permit it). These recommendations take into account the user's preferences, which increases the probability of follow through for the given suggestion. This is calculated based on past events similar to the recommendations with a positive emotional response. 
% \item 
\textbf{Goals/Fear} motivate the events which the user doesn't necessarily like but are important for achieving a long term goal or to avoid an outcome. Waiting in traffic while commuting to work is an example of this type of motivation. These events can be determined from frequent patterns in event history and include events which are repeated even regardless of emotional response.
% \item 
\textbf{Social/External pressure} motivate the events where external pressure from other people maybe a factor. This factor controls significant portions of social behavior and hence may influence the user to pursue activities which they usually would not pursue. We use social media connections along with social media activity to understand which events are influenced by other people or social groups.
% \end{itemize}

\item[Ability] represents the accessibility in performing the suggested behavior. This factor is an interplay of the individual's surroundings and their intrinsic capability. When ranking the recommendations, we will integrate the information about environment from Eventshop (as specified in EnviroSense) and the event history of the person. This will let us know whether the person's ability and the environmental constraints required for the event are satisfied or not, and matches their needs to the available resources. For example, we would only suggest driving to get a healthy lunch if Bruce had access to a car (intrinsic capability), and the traffic conditions (surrounding resources) allowed him to travel in time for his next meeting.

\item[Triggers] are reminders or suggestions personalized to the individual's needs to accomplish the task\cite{Fogg2009}. Two broad categories of triggers include: 1. Natural triggers from physiological events occurring in the body, such as hunger. 2. Synthetic triggers are provided by the system and aimed at facilitating the occurrence of an event. These can be in form of a notification on a smartphone or an intervention from a friend or relative. Synthetic triggers are synergistic when coupled with relevant natural triggers. For example, a notification about healthy food options when the person is hungry. In our system, synthetic triggers in the form of parsed dish menus are used to enhance the person's ability by recommending items which are similar to the person's preferred activities but have a positive impact on the user's health (in our case, food items similar to what user likes but comparatively healthier). This helps us match the user's needs to the available resources and generate recommendations which meet the criteria of the behavior model. Bruce's lunch menu on HB follows from this analysis (Figure \ref{fig:screens}).
\end{description}

The variance in recommendations are calibrated based on the different types of events in the user's history. HB caters to the range of events that hover in the vicinity of user event history. This increases the effectiveness of recommendations.

\subsection{E. Prevention}
By preventing negative outcomes, we accomplish several tasks. Users are given direct feedback to understand actions they are taking are working to benefit themselves. This keeps user motivation up and encourages further participation in ownership of their own health. Prevention also depends on informing the individual of risks regarding their choices. In the USA, calories are printed on menus of large franchise restaurants by law. This allows consumers to have direct basic information on what they are consuming. Similarly, the practice of having cancer warning signs on cigarette packages or labeling alcohol warnings for pregnant women are designed to inform the consumer of their choice. HB and any other system derived from the P5C concept focus on informing the user in a personalized fashion.

Preventing problems and managing health information benefits quality of life. Morbidity attached to disease management is a large factor or reduced quality of life. Giving patients the ability to prevent the progression of their disease has been shown to improve quality of life \cite{uczynski2016EmpowermentObesity}. As a bonus, corrective actions through lifestyle factors positively affect multiple comorbidities at the same time. These interventions reduce diabetes, dyslipidemia, and prevent cardiovascular disease \cite{Khavandi2017,Moon2017PreventionMellitus, Khavandi2017}.
Prevention involves early detection from continuous monitoring. Most importantly, earlier diagnosis makes treatment through lifestyle interventions much more effective. Continuous monitoring also allows the medical team to modify treatment more appropriately through tighter coordination. Ultimately, the prevention of deteriorating health conditions keeps the individual in the steady state of optimal health. This is the original goal of cybernetic systems, which aligns perfectly with the goals of an optimal health system.

\section{Current Status}

\begin{figure}[t!]
\centering
\includegraphics[width=0.8\columnwidth]{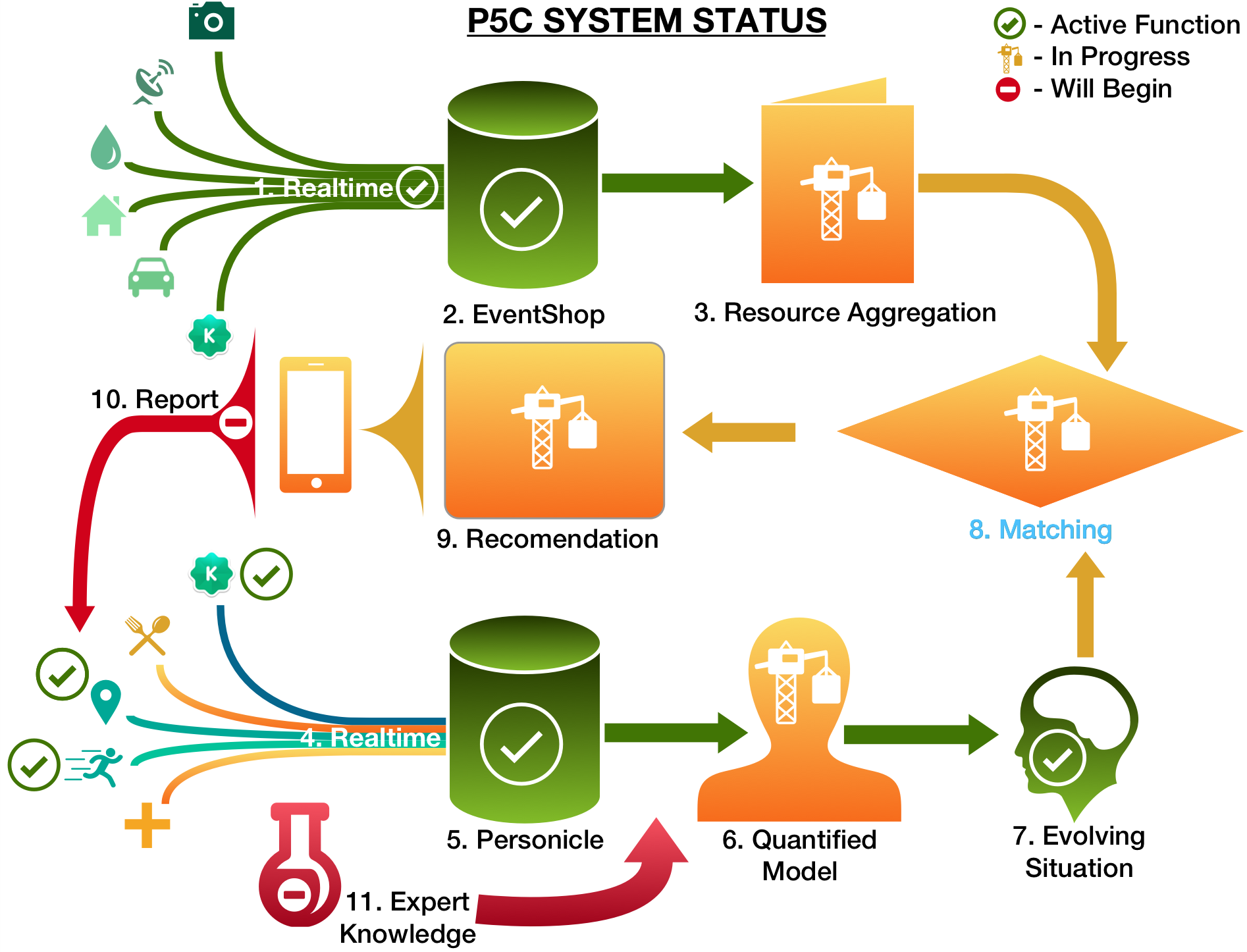}
\caption{A visualization of our progress to build P5C.}
~\label{fig:status}
\end{figure}

At the time of writing this project we have several modules fully working, some under active construction, and a couple that will begin construction soon (Figure \ref{fig:status}). Our functional active closed-loop system will be ultimately deployed in the hospital setting. The construction status for the system modules follows from the integration as follows.

\begin{enumerate}
\item Realtime data: Heterogeneous data sources from real-world events, such as social sources (e.g. Twitter, Facebook and Flickr), environmental sources (e.g. flood, hurricane, asthma, flu, population, pollution, and weather), camera, and traffic etc \cite{gao2012eventshop, pongpaichet2013eventshop}.
\item Eventshop: Providing operators for data stream ingestion, visualization, integration, situation characterization, and sending out alerts \cite{singh2016situation}.
\item Resource aggregation: Situation recognition obtaining actionable insights from observed spatio-temporal data \cite{singh2016situation, gao2012eventshop, pongpaichet2013eventshop, tang2015geospatial, tang2016integration}.
\item Realtime data: Heterogeneous data sources from human-related sensor data by smartphone and wearable sensors (e.g. activity, step, GPS, venue, call, calendar, wifi connection, smartphone application, photo, ambient light, ambient sound etc.) \cite{jalali2014personicle,oh2015intelligent,JordanThesis:2015}.
\item Personicle: Identifying semantic-level life events using heterogeneous data sources and creating a chronicle of life events \cite{jalali2014personicle,JordanThesis:2015}.
\item Quantified model: Comprehensive human model using objective quality data from heterogeneous data sources for individuals \cite{jain2014objective}.
\item Evolving situation: Human behavior analysis with causal modeling across multimedia data streams \cite{jalali2015bringing, jalali2016interactive, jalali2016human, jalali2016framework}.
\item Matching: Merging environmental situation and personal situation \cite{tang2015habits, tang2016research}.
\item Recommendation: We are actively engaged in developing a recommendation engine.
\item Reporting and 11. Expert Knowledge: We will begin integrating this aspect soon.
\end{enumerate}

\section{Conclusions and Future Challenges}
Cybernetic principles lay the foundation of building health systems that are responsive in keeping individuals at optimal health over the course of a lifetime. Our conceptual foundation of P5C is constructed with the most fundamental principles in control theory while incorporating the ability to seamlessly integrate both present and future technological advancements. This is an absolute necessity to transform the archaic practices of current day health care, especially in light of how many diseases complications are preventable. The most important concept we describe is how these parts integrate to form a true closed-loop system. These closed-loop systems have been refined in mechanical systems over decades and in biological systems for millions of years. 
This also ties together the span of our health ecosystem to work in synergy. It is absolutely necessary to use these principles to bridge the virtual and real world together to move human health forward.

Each individual block of P5C (personalized, predictive, precision, persuasion, prevention) for HB is under active development to be completed for beta testing in patients by March 2017. Real health progress relies on an interdisciplinary effort between hospital clinicians, engineers, computer scientists, and bioscience researchers. Our work focuses on the translation of interdisciplinary academic progress into real systems that patients will benefit from. HB is just one incarnation of P5C in a hospital focused application that we are launching with the UCI Health Diabetes Center and a cornerstone of the interdisciplinary UCI Institute for Future Health. There will be various technical challenges during large scale deployment of any P5C system. Consideration of how various countries have different habit patterns, regulations, medical systems, sensor and network connectivity, and environments is essential. 
Tackling the reliability of sensor data is essential for systems like this to properly function. Integration of natural language processing for medical literature will also improve the ability to quickly disseminate actionable information to the masses. Security and privacy are of utmost concern in all circumstances for patients and providers. These challenges are constantly being addressed as P5C systems grow.

\section{Acknowledgments}
We thank Jonathan Lam for his design and user interface contributions. This research is partially funded by the National Institute of Health (NIH, United States of America) as part of the Medical Scientist Training Program (MSTP) and the Cardiovascular Applied Research and Entrepreneurship (CARE) grant under \#T32GM008620-15. Additionally funding is provided by the UC Irvine Donald Bren School of Information and Computer Science.

% BALANCE COLUMNS
\balance{}

% REFERENCES FORMAT
% References must be the same font size as other body text.
\bibliographystyle{SIGCHI-Reference-Format}
\bibliography{sample}

\end{document}